# Novel Calibration Method for Switched Capacitor Arrays Enables Time Measurements with Sub-Picosecond Resolution

D. A. Stricker-Shaver, S. Ritt and B. J. Pichler

*Abstract*—**Switched capacitor arrays (SCA) ASICs are becoming more and more popular for the readout of detector signals, since the sampling frequency of typically several gigasamples per second allows excellent pile-up rejection and time measurements. They suffer however from the fact that their sampling bins are not equidistant in time, given by limitations of the chip process. In the past, this limited time measurements of optimal signals to standard deviations (σ) of about 4-25 ps in accuracy for the split pulse test, depending on the specific chip. This paper introduces a novel time calibration, which determines the true sampling speed of an SCA. Additionally, for two independently running SCA chips, the achieved time resolution improved to less than 3 ps (σ) independently from the delay for the split pulse test, when simply applying a linear interpolation. When using a more advanced analyzing technique for the split pulse test with a single SCA, this limit is pushed below 1 ps (σ) for delays up to 8 ns. Various test measurements with different boards based on the DRS4 ASIC indicate that the new calibration is stable over time but not over larger temperature variations.**

## I. INTRODUCTION

SWICHED CAPACITOR ARRAYS (SCA) are application specific integrated circuits (ASIC) which allow transient analog signal recording at high sampling speeds. They have been first used for particle detector readout back in the 1980's as shown by Kleinfelder [1]. The principle is to use an array of capacitors in sample-and-hold mode. A fast sequence of write pulses allows the recording of analog waveforms in these capacitors, which can later be read out and digitized at a much lower speed. While early chips used shift registers to generate the write pulses, it was soon realized that using inverter chains as delay lines boost the sampling speed into the gigasamples per second (GSPS) region as demonstrated in [2]. Following the advances in CMOS technology, SCAs have become faster over the years, and current chip versions from different groups allow sampling speeds in the rage of 2-15 GSPS [3][4][5][6].

They have signal-to-noise ratios (SNR) of 10-13 bits equivalent and sampling depths of 256 to 64k cells. Paired with typical power consumptions of 10 mW per channel, these chips are excellent alternatives for commercial flash-ADCs. They all share however the disadvantage that the time required to read out the capacitor cells causes dead time. Depending on the number of channels and cells to read out, this dead time is in the typical rage of 1-100 micro seconds, which limits the application to cases where one can use a trigger to limit the number of events to typically 100-1000 events/s. The field of application for SCAs therefore lies in areas where one has a low trigger rate, but requires excellent time resolution and pile-up rejection. This includes for example particle physics at the intensity frontier[7], Cherenkov telescopes in gamma-ray astronomy[8], time-of-flight (TOF) applications [9] and neutrino physics[10]. Also, in medical imaging, specifically in positron emission tomography (PET), where TOF plays an important role[11], SCAs are a candidate for future applications. In [12] a coincident resolving time (CRT) of 100 ps (FWHM) was demonstrated for PET signals using LaBr$_3$ as a scintillator. This represents a single time resolution of ~30 ps (σ) and requires precise electronics to measure this time.

Several TOF measurements were made recently with various SCA chips. In the readout of micro channel plates, a single detector resolution of ~15 ps (σ) was achieved, which is comparable with the best commercial combination of constant-fraction-discriminators with time-to-amplitude converters, but at a much lower cost per channel [13]. The readout of straw tubes [14] via a time measurement gave good results, although the accuracy was limited by the imperfect time calibration (TC) method used for the SCA chip.

Over many years, the general opinion was that time measurements with SCA chips are limited to about 4-25 ps (σ) [6], caused by the time jitter inside the chips. The authors have found however that SCA chips can perform much better if the correct TC is applied. This allows pushing the achievable time resolution by about one order of magnitude to below one picosecond.

## II. THEORY OF OPERATION

Most modern SCAs use a kind of inverter chain to generate the write pulses which open analog switches to sample the input signal. Fig. 1 shows a simplified schematics of the DRS4 chip, which is the fourth generation in a family of SCAs

This work was supported in part by the Deutsche Forschungsgemeinschaft (DFG) Grant PI771/3-1, the Schweizer Nationalfonds (SNF) Grant 200021_137738 and the Swiss Werner Siemens Foundation.

D. A. Stricker-Shaver and B. J. Pichler are with Werner Siemens Imaging Center, University of Tübingen, Tübingen, Germany.

S. Ritt is with Paul Scherrer Institute (PSI), Villigen, Switzerland (e-mail: stefan.ritt@psi.ch).

developed at PSI [5]. The advantage of this technique is that a simple inverter chain used as a tapped delay line can run much faster than any shift register, easily reaching 10-20 GSPS with modern chip technologies. The disadvantage however is that the transition time of an inverter depends on parameters like temperature and supply voltage. To address this problem, most SCAs use a delay-locked loop (DLL) or a phase locked loop (PLL) to stabilize the sampling frequency to an external constant reference clock.

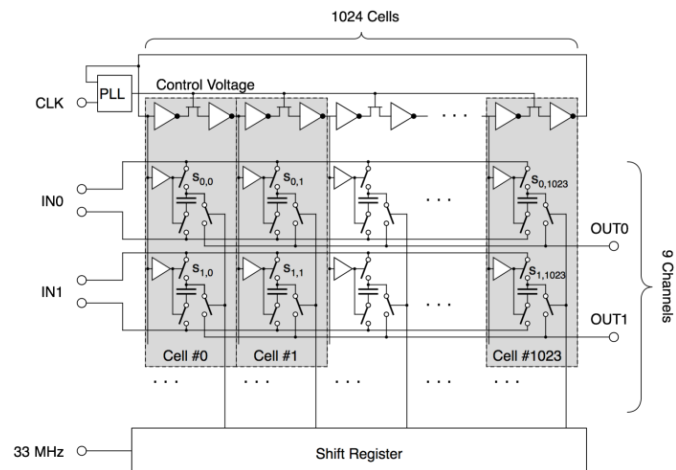

Fig. 1: Simplified schematics of the DRS4 chip.

Furthermore, mismatch between transistors in the CMOS process causes each inverter to have different but fixed transition times even if the other parameters are kept constant. Since this effect comes from the actual geometrical size of a transistor and its doping properties, it is stable over time and can be corrected for by measuring the transition time of each inverter and correcting for it. This measurement and its correction is the main topic of this paper and will be detailed in chapter IV.B.

III.    LIMITATIONS OF TIME MEASUREMENTS

After a proper TC, the accuracy of a time measurement with an SCA chip is limited by the residual random jitter of the transition time of the inverter chain. Each inverter has a voltage threshold at the input. Crossing this threshold turns the inverter high or low. While this transition is very stable, any noise on the input signal will cause a time jitter of the inverter. Careful chip design allows minimizing this noise, causing modern SCA chips to have typical inverter chain time jitters below 1 ps.

To measure the arrival time of a certain electrical pulse e.g. in particle physics, the pulse time is typically extracted from the first rising or falling edge of the waveform, depending on the pulse polarity. The simplest case is to use a single threshold discriminator. In the case of waveform digitizing, the equivalent can be achieved in the digital domain by comparing the digitized voltage of the sampling points with a fixed value. To achieve a time resolution exceeding the sampling interval, adjacent samples can be interpolated linearly or with cubic functions. In Fig. 2a an interpolated line from an ideal signal intersects a given threshold at time $t_1$ depicted by an open square. Any voltage noise on the measured signal causes time jitter as shown in Fig. 2b. If some voltage noise "raises" the signal by an amount $\Delta u$, the linearly interpolated line intersects the same threshold at a time $t_1'$ different from $t_1$, as depicted by the grey square. From Fig. 2b one can easily derive the formula for the time accuracy $\Delta t$ as

$$\frac{\Delta u}{\Delta t} = \frac{U}{t_r},$$   (1)

where $U$ is the signal height, $t_r$ the rise time and $\Delta u$ the voltage noise as shown in Fig. 2.

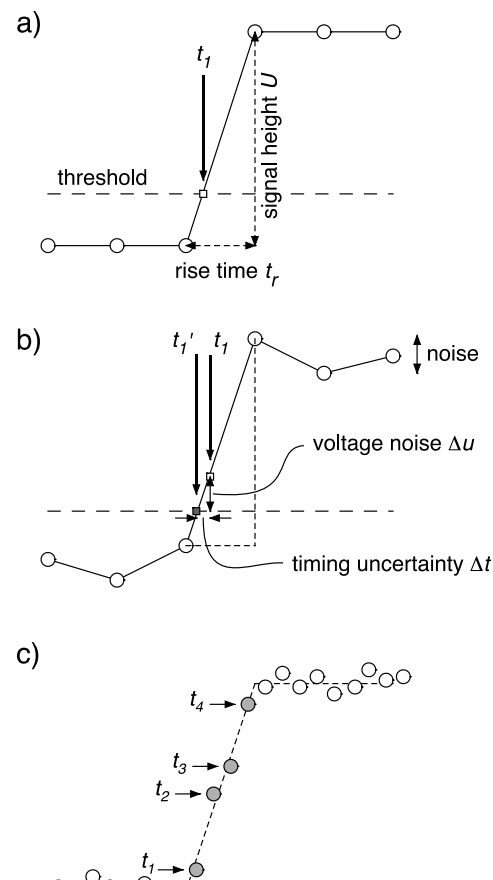

Fig. 2: Time estimations for a leading edge in the ideal case (a), in the presence of noise (b) and for several sampling points lying on the edge (c).

The time resolution can be improved by sampling the signal at a higher frequency. Fig. 2c shows the same signal sampled with four times higher sampling rate. The sampled points scatter around the signal indicated by the dashed line. Now several points lie on the signal edge, shown as grey circles. If the voltage noise of these points is statistically independent (as it is the case e.g. for ADC quantization noise), each point allows a separate measurement of the edge time, and thus reduces the time uncertainty of the edge by $\sqrt{n}$ where $n$ is the number of points lying on the edge. The value of $n$ is also determined by the product of sampling frequency $f_s$ and the signal rise time $t_r$. This gives the theoretical time accuracy in dependence of the SNR which can be expressed as $U/\Delta u$, the sampling frequency $f_s$ and the signal rise time $t_r$. Adopting this to (1) and solving for $\Delta t$ gives

$$\Delta t = \frac{\Delta u}{U} \cdot t_r \cdot \frac{1}{\sqrt{n}} = \frac{\Delta u}{U} \cdot \frac{t_r}{\sqrt{t_r \cdot f_s}} = \frac{\Delta u}{U} \cdot \frac{\sqrt{t_r}}{\sqrt{f_s}} \quad . \tag{2}$$

From (2) it becomes clear that not only a high sampling frequency is important for a precise time measurement, but also the SNR and the signal bandwidth. It should be noted that (2) is only a simplified formula, since the actual resolution depends on the exact waveform shape and the noise spectrum. It is however a good estimator which has been verified by the authors and other groups[15].

For very fast detector signals, such as pulses from microchannel plates (MCP) which can have $t_r$ below 100 ps, the sampling gets limited by the bandwidth of the signal chain. The limiting factor can be a pre-amplifier, a long cable or the SCA chip itself, which is further detailed in [4]. The -3dB bandwidth $f_{3dB}$ determines the signal rise time seen by the SCA as

$$t_r \cong \frac{1}{3 f_{3dB}}. \tag{3}$$

Putting this into (2) results in the time accuracy in dependence of the bandwidth $f_{3dB}$ as

$$\Delta t = \frac{\Delta u}{U} \cdot \frac{\sqrt{t_r}}{\sqrt{f_s}} = \frac{\Delta u}{U} \cdot \frac{1}{\sqrt{3 f_s \cdot f_{3dB}}}. \tag{4}$$

| Case | $U$ (mV) | $\Delta u$ (mV) | $t_r$ (ns) | $f_{3dB}$ (MHz) | $f_s$ (GSPS) | $\Delta t$ (ps) |
|------|------|------|------|------|------|------|
| a) | 10 | 1 | 1 | 333 | 5 | 45 |
| b) | 450 | 1 | 1 | 333 | 5 | 1 |
| c) | 100 | 1 | 0.35 | 950 | 5 | 2.6 |
| d) | 500 | 0.35 | 1.6 | - | 5 | 0.5 |
| e) | 63 | 0.35 | 0.2 | - | 5 | 1.1 |

Table 1: Theoretical limit of the achievable time resolution $\Delta t$ for certain signal and sampling parameters.

Table 1 lists various scenarios, which are relevant for this paper. Cases a) – c) assume a typical environmental or preamplifier noise $\Delta u$ of 1 mV. The SCA internal noise in case of the DRS4 is 0.35 mV (case d) and e) ). In all five cases the SCA is running at a sampling speed of 5 GSPS. Assuming a rise time of 1 ns, which is typical for many photomultipliers, it becomes clear that small signals in the 10 mV range will never give a time accuracy better than a few ten ps. Only a significantly higher signal – for example by means of a low-noise, fast preamplifier – can bring the resolution into the picosecond range. If one uses even faster pulses, the bandwidth gets limited by the SCA itself at some point. In the case of the DRS4 chip this bandwidth is 950 MHz, which limits the time resolution for a SNR of 100:1 to 2.6 ps (case c). Case d) and e) are relevant for the TC in chapter IV.B., which is performed with a 100 MHz sine wave with an amplitude of about 1V.

## IV. MATERIALS AND METHODS

The TC methods were evaluated with the DRS4 Evaluation Board version 3 (board A) with 12 bits, 5 GS/s and 4 channels provided by PSI[16] and the DRS4-chip based V1742 (board B) with 12 bits, 5 GS/s and 32 channels, provided by CAEN, Italy[17].

We further compared the performance results of board A & B with a LeCroy Wave Runner 6050A Oscilloscope (board C) with 8 bits, 5 GS/s and 4 channels and with the V1751 (board D) with 2 GS/s, 10 bits and 4 channels provided by CAEN. In this paper the presented measurements of board A, B and C were taken at a sampling speed of 5 GSPS, while board D used a sampling speed of 2 GSPS.

### A. Voltage Calibration

Since voltage errors and time errors on an SCA chip are correlated, it is important to correct for any voltage error before a TC can be done. It consists of three corrections.
Firstly, the voltages of the stored waveform show slightly different offsets and gains for each sampling cell. In the DRS4 chip, this comes mainly from the fact that each sampling capacitor is read out by a separate buffer, which has a typical random offset of 10-20 mV. These offsets can be measured by connecting the input to a DC voltage, e.g. 0 V, and then subtracting these offsets in each measurement as an offset correction.

Secondly, a time-dependent readout offset correction is performed. This compensates for small supply voltage variations when the DRS4 chip is switched from sampling to readout mode. The different power consumptions of the DRS4 chip in these two modes causes a small dip in the power supply voltage, which cannot be recovered completely by the linear regulator or the blocking capacitors. The dip causes the DRS4 output to shift by about 2 mV for about 10 μs after it has been stopped.

Thirdly, the gain correction for each cell is done by applying a DC voltage of 800 mV to the input and measuring the response of the cell. The third correction is performed for board A but not for board B.

It is important to do  the offset correction at the same voltage level as used later for time measurements, because the gain does not show an absolutely linear behaviour. An example is given in Fig. 14, where for board B an external bias offset gets applied to shift the input range of the DRS4 chip after the voltage calibration.

Some SCA chips have the problem that some residual imprint of the previously stored signal distorts the last sampled waveform. This can cause the chip to respond differently to DC signals and to AC or transient signals. In the case of the DRS3 chip, this so-called "ghost pulses" could amount to 2-5 percent distortion of a waveform, depending on the sampling speed. This problem has been fixed for the DRS4 chip by issuing a clear cycle before a storage cell is written. Both sides of the storage capacitor are connected to ground by additional analog switches for a few nano seconds before every write cycle, thus removing efficiently any previously stored charge in the cell.

## B. Time Calibration (TC)

The sampling intervals of the DRS4 chip are not equidistant, but constant over time. This means the DRS4 has to be calibrated before a precise time measurement can be made. One can find such TC methods, e.g. in [6],[9],[18] and [16]. In this paper the new TC will be referred to as TC N. The time calibration from board A will be named TC A [16].

Consistently, the TC from board B will be called TC B. The TC B is unknown, since CAEN ships its boards already calibrated and nothing is mentioned in the manual.

In the following we will use $\Delta t_i$ as the effective sampling interval between cell# $i$ and cell# $i+1$. We define $\Delta t_i$ as the time difference between the opening of the analogue switches $S_{0,i}$ and time point $S_{0,i+1}$ as illustrated in Fig. 1. From this follows that the integrated or "global" time difference between cell# $k$ and cell# $q$ is given by:

$$\Delta t_{k,q} = \sum_{i=k}^{(q-1)} \Delta t_i . \tag{5}$$

We digitized a known sine wave to perform TC N. The frequency of this sine wave $f_{TC\,N}$ should be adjusted accordingly to the sampling speed range of the SCA. For the DRS4 sampling speed range of 0.7 - 5 GSPS , we recommend the frequency of 35 MHz in order to achieve best results. 35 MHz is the highest possible frequency due to (6) for the critical 0.7 GSPS case. If the sine frequency is much higher, the linear interpolation method would break down, if the frequency is lower, the duration of the TC would increase. Thus, TC N works in the frequency ranges

$$f_{TC\,N} \in \left[ \frac{2}{n} \cdot f_{SCA} \quad , \quad \frac{1}{20} \cdot f_{SCA} \right], \tag{6}$$

where $f_{SCA}$ is the nominal sampling frequency of the SCA and $n$ stands for the number of it cells.

In the following sections we always run the DRS4 at a sampling speed of 5 GSPS. For TC N we used the highest frequency we had available to reduce calibration time. We therefore used a sine wave of 100 MHz with an 1 V peak-to-peak amplitude, which is the middle range of (6).

TC N consists of two parts. The first part estimates the effective sampling intervals $\Delta t_i$ by measuring voltage differences between two neighboring cells and is called "local" TC. The second part refines the sampling intervals by measuring time differences between cells that are far apart and is therefore called "global" TC. While the local TC effectively corrects the differential time non-linearity, the global TC reduces the integral time non-lineariy as will be shown later in Fig. 5.

### 1) Local TC

Two publications ([9],[19]) have already introduced the basic idea that we call the local TC. We devolved this idea independently and combining it with our global TC described later, which would improve the current results significantly as in Fig. 5.

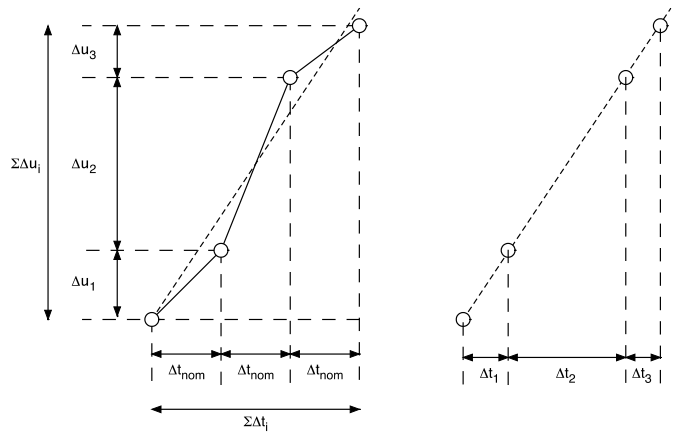

Fig. 3: The correlation between voltage differences $\Delta u_i$ and time differences $\Delta t_i$ of a rising edge can be used for the local TC of an SCA chip.

The approach is that $\Delta t_i$ is proportional to the measured voltage difference between neighboring cells when applying a linear increasing or decreasing signal, such as a saw-tooth waveform for example. The intercept theorem results in (7) and is illustrated in Fig. 3:

$$\frac{\Delta t_i}{\Delta U_i} = \frac{\sum \Delta t_i}{\sum \Delta U_i}, \tag{7}$$

where $\Delta U_i$ is the voltage difference between cells $i$ and $i+1$. For an SCA with $n$ cells we know

$$\sum_{i=1}^{n} \Delta t_i = \frac{n}{f_{SCA}} . \tag{8}$$

When combining (7) and (8), we can calculate all $n$ time intervals $\Delta t_i$ as:

$$\Delta t_i = \frac{\Delta U_i \cdot \frac{n}{f_{SCA}}}{\sum \Delta U_i} . \tag{9}$$

The exact sampling speed of the SCA that might deviate from $f_{SCA}$ is not required since it will be determined in the global TC afterwards.

Using rising and falling edges of the TC signal will result in two calibrations. Averaging over these two calibrations will cancel any residual voltage offset:

$$\Delta t_i = \left( \Delta t_{i,falling} + \Delta t_{i,rising} \right) \cdot \frac{1}{2}, \tag{10}$$

where $\Delta t_{i,falling}$ and $\Delta t_{i,rising}$ stand for the time differences calculated by the falling and rising edges, respectively. Tests have indicated that using the rising and falling edges of a sine wave like the one in Fig. 4 are good enough to obtain an acceptable local TC. In a digitized waveform only the $\Delta t_i$ for cells on the slopes of the sine wave can be determined. The procedure therefore has to be repeated for several sine waves with a random phase relative to the SCA clock. 1000 digitized sine waves give a decent result for the local TC when simply using the arithmetic mean values.

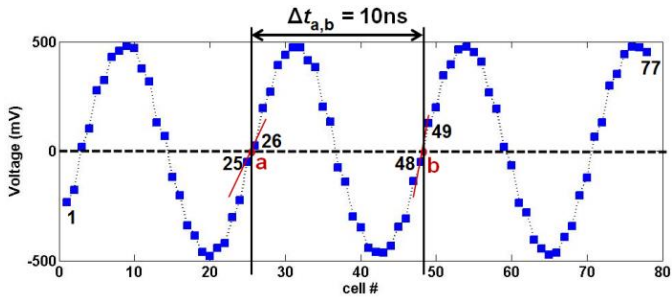

Fig. 4: First 77 cells of the 1024 cell array of a DRS4 sampling a 100 MHz sine wave at a sampling speed of 2.5 GSPS. This signal is used for the local TC and the global TC.

For the 5GSPS case, the local TC is using sampling points in the range of ± 100 mV.

### 2) Global TC

The global TC measures one or more periods of the 100 MHz sine wave. The period is determined by linearly interpolating sampling points below and above zero volts and measuring the time between the intersections of the interpolated lines with the zero line as shown in Fig. 4. In this example the zero crossing is between cells #25 and #26 and between cells #48 and #49, leading to two corrections

$$t_{cor} = \frac{U_k}{\Delta U_{k-1}} \Delta t_{k-1} - \frac{U_q}{\Delta U_q} \Delta t_q \ , \qquad (11)$$

which constrain the time between the cell# $k = 26$ and cell# $q = 48$ using the known period time of $f_{TC\,N}$ as in

$$m \cdot \frac{1}{f_{TC\,N}} \stackrel{\text{def}}{=} t_{k,q} + t_{cor} \quad , \qquad (12)$$

where $m$ stands for the factor describing the multiples of the measured periods. $U_i$ is the voltage measured at cell# $i$.

In Fig. 4 the global time difference $\Delta t_{a,b}$ has to be $1 \cdot 10$ ns. The two points $a$ and $b$ are artificial points and stand for the zero crossings of two rising edges of the digitized sine wave.

Finally the global TC is computed iteratively by correcting the local TC each time we measure a multiple of the period time $m \cdot 1/f_{TC\,N}$.

$$
\begin{aligned}
\Delta t_{k,new} &= \Delta t_k \cdot u_{cor} \\
\Delta t_{k+1,new} &= \Delta t_{k+1} \cdot u_{cor} \\
&\vdots \\
\Delta t_{q-1,new} &= \Delta t_{q-1} \cdot u_{cor}
\end{aligned}
\quad \text{with} \quad u_{cor} = \frac{m}{f_{TC\,N} \cdot (t_{k,q} + t_{cor})} \ ,
$$

$$\qquad (13)$$

where $\Delta t_{k,new} \dots \Delta t_{q-1,new}$ stand for the corrected effective sampling interval. $\Delta t_k \dots \Delta t_{q-1}$ stand for the old data set of effective sampling intervals that are going to be corrected. The first iteration will start with the data set provided by the local TC (9). In the following iterations the $\Delta t_{i,new}$ provided by (13) will be retitled in $\Delta t_i$ in order to apply (13) again every time a new $t_{k,q}$ is measured. Iterating over typically 1000 digitized sine waves is usually enough to obtain a precise global TC. Ideally one should treat falling and rising edges separately and determine the global TC after applying (10). For practical reasons, the measured sine waves in the local TC can be "recycled" for the global TC. In Table 1 case e) one can see that the expected time resolution is about 1.1 ps for a single interpolation. In this case the voltage difference between sampling points of a rising edge is in average about 63 mV with an average sampling interval of 0.2 ps.

The error arising from the linear interpolation of the sine wave is predictable and can be estimated. Simulations for an expected sampling interval of 0.2 ns show that the global TC error is less than 0.08 ps, and less than 0.7 ps in the worst case scenario, where some sampling intervals could vary up to 0.4 ns. Increasing the TC frequency will increase the maximum error of the global TC.

The global TC improves the local TC for two reasons. First, the local TC is never perfect. Any measurement has a statistical error, which accumulates if one integrates over many measurements. Second, the SCA cells have different effective analog bandwidths along the chip. The cells close to the input pin see a smaller resistance of the signal bus inside the chip than the cells far away from the input pin. This causes slightly different rise times for the calibration sine wave in different cells, and causes a systematic error for the local TC, which is then removed efficiently by the global TC.

### C. Board Time Resolution Tests

To examine the time resolution of the four boards and the quality of TC N, 4 different performance tests were applied. The time resolutions for all performance tests are given as standard deviation (σ) and refer to resolution on time difference. All these σ values were extracted from a Gaussian fit applied to their distributions. The σ coming from the fit is insensitive to appearing outliers, which are typically less than 0.1 % of all measurements. The value of a calculated standard deviation (often the terminology root-mean-square (RMS) is used) compared to σ received from the fit was never more than 0.2 ps increased. An example is given in the section Results in Fig. 13.

### 1) The Period Time Test (PT-test)

The 100 MHz sine wave used for the TC N is also used for the PT-test. Fig. 5 shows the measured period time between two zero crossings of the rising edges, which should be equal to 10 ns. The period time is plotted over the cell number that is left of the first zero crossing. Thus, $\Delta t_{a,b}$ from Fig. 4 would be represented by cell# 25 in Fig. 5. The top plot shows the time for an uncalibrated DRS4 chip, the middle after the local TC and the bottom after the global TC. While the local TC effectively corrects variations between neighboring cells, a residual inaccuracy with more global structures is left over, which can be seen in the middle plot. The global TC then corrects these global structures, which leads to a flat distribution shown in the bottom plot. This method is therefore very powerful to determine the accuracy of a TC.

In Table 2 we compare the distribution of the period time (i.e. the projection of the distribution in Fig. 5 bottom onto the Y-axis) for different hardware. For each measurement, 10000 events were digitized and averaged. When digitizing the 100 MHz sine wave with 5 GSPS, we measured average voltage difference between two consecutive samples of $\Delta U_i$= 63 mV around the zero crossings. This effective signal height was used to predict the result of the SP-test by using (1).

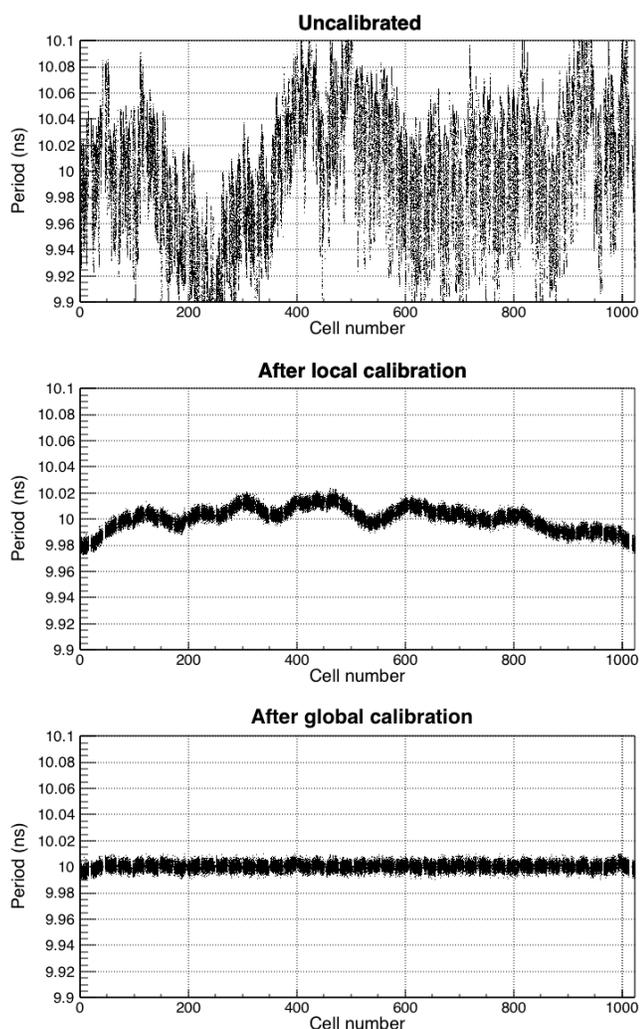

Fig. 5: Effect of the local TC and the global TC when used to determine the period of a 100 MHz sine wave (PT-test).

### 2) The Split Pulse Test (SP-test)

We evaluated the actual time precision for the tested boards as shown in Fig. 6a with the SP-test. A short pulse is generated by a function generator. The signal is split and the time difference between the arrivals of the two signals is measured. We changed the arrival time of one of the signals by applying a variable cable delay (Fig. 6a and Fig. 7).

In most cases we used a simple Digital Leading Edge (DLE)-discriminator to evaluate the time resolution. This means, we interpolated between two points of the rising edge to obtain the time information at a given threshold. A global threshold (TH) of 300 mV was used for all SP-tests (Fig. 7). In Fig. 10 and Fig. 14 the THs were changed individually for a single SP-test measurement (50 ns cable delay) resulting in a time resolution matrix. One can see that the global TH of 300 mV is a good compromise for all SP-test that use a DLE-discriminator.

The time resolution improves by using more points. Thus, we also performed the SP-test with a DLE-discriminator by fitting though 6 points of the rising edge.

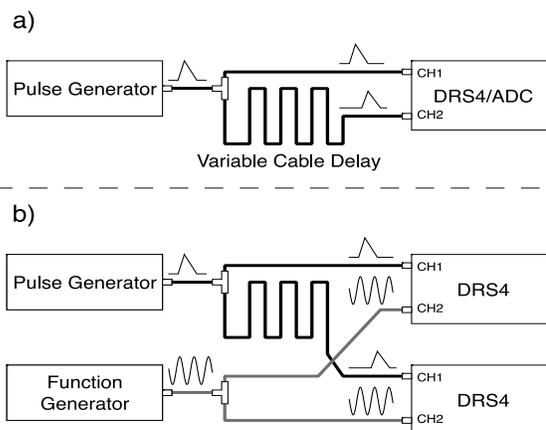

Fig. 6: Layout of the split signal time measurements for two channels of one chip (a) and between two chips synchronized by a sine wave (b).

Best results for the SP-test were achieved by cross-correlating both channels. In this case all non-baseline related sampling points of the pulse were used. Thus, we used 50 points of the split signals. Before, 50 linear interpolations were made for each channel in order to result in equidistant sampling points.

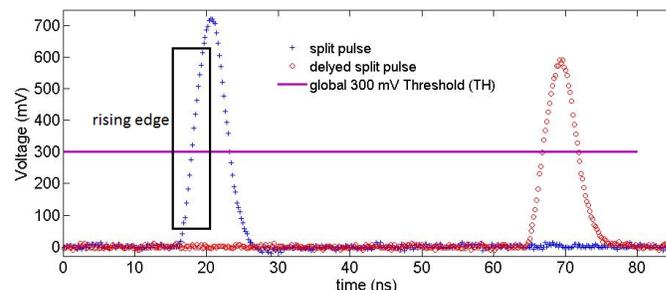

Fig. 7: Split pulse digitized at 5 GSPS for the SP-test with board C and a full range of ~1.1V. The left signal height peaks around 720 mV and the delayed signal peaks at around 610 mV due to cable attenuation. The time difference between the signals is approximately 50 ns. The split pulse rise time is around 3 ns or 15 sampling points.

Also the time resolution ($\sigma$) between two independently running DRS4 chips was tested (Fig. 6b). In this case a 100 MHz clock was split and additionally sampled in a separate channel in each of the chips. The measured phase shift of the split clock was used to synchronize the two DRS4 chips.

### 3) The Coincident Resolving Time Test (CRT-test)

Fig. 8 shows the setup of a PET time resolution measurement (CRT) when using the DRS4 chip with and without applying TC N instead of TC B. We used two 5mm · 5mm · 5mm LSO:Ca crystals as scintillators glued to two fast PMTs (Hamamatsu R10560). The Results are displayed in a matrix form, where the time resolution ($\sigma$) is given as a function of the thresholds for both PMT channels. The thresholds are measured in % of the averaged 511keV photo-peak height of the individual channel.

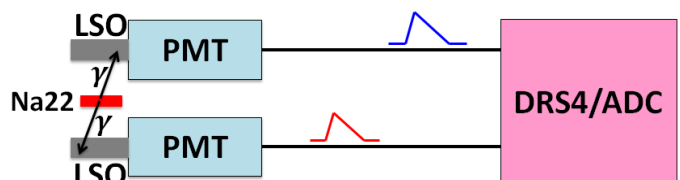

Fig. 8: Layout for the PET time measurement.

### 4) The Temperature Time Dependence Test (TTD-test)

For the TTD-test board A was used. When performing the TC N, 10 global time differences $\Delta t_{1,100}$ to $\Delta t_{900,1000}$ were measured at 8 different temperature levels in 5 °C steps from 5 °C to 40 °C. Aftereach temperature change, board A had to run 12 hours in a temperature controlled box before the TTD-test was performed. The temperature was measured with two sensors, one located in the box and the one included in board A. The temperature on the board was larger in average by 12°C compared to the box temperature due to the self-heating of the board.

## V. RESULTS

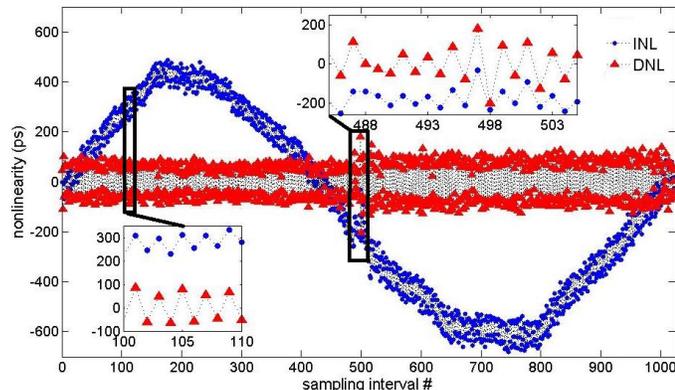

Fig. 9: Integrated nonlinearity (INL) and differential nonlinearity (DNL) of all sampling intervals measured by TC N for channel# 4 of board A running at 5 GSPS.

An alternating sampling behavior was expected when looking at the slopes of the waveform in Fig. 4. The sampling intervals of adjacent cells differ alternating by about 70 ps for the DRS4 from the nominal width of 200 ps at 5 GSPS, which can be seen in Fig. 9. The integrated nonlinearity (INL) varies in this case from -680 ps to 486 ps. The differential nonlinearity (DNL) is alternating stronger and has extremes at interval# 497 = 170 ps and interval# 498 = −203 ps. The DNL smaller than -200 ps means that the signal reaches cell# 499 always 3 ps before it gets sampled in cell# 498 which can be explained by the layout of the chip. The input bus is routed due to some constraints such that the signal reaches cell #499 before it reaches cell #498. From 20 tested DRS4 chips, cell# 498 was the only cell showing this behavior and is typically in the DNL range of -205 ps and -150 ps when sampling at 5 GSPS.

By using the TC N we also verified that board B is running at the nominal 5 GSPS. We evaluated that all previous versions of the DRS4 evaluation boards version 5, such as board A, were running at 5.1206 GSPS instead of the expected 5.1200 GSPS.

### 1) The Period Time Test (PT-test)

Table 2 shows among others the results for the oscilloscope (board C). The resolution can be improved by increasing the gain, but with the drawback of a clipped signal. The full 100 MHz 1 V signal was only visible in the first case on the oscilloscope screen and results for channel# 2 in a time resolution of 16.7 ps (σ). By increasing the gain the information of the full waveform got lost but increased the

SNR and therefore increased the time resolution. The best gain uses the 8 bits for a full range of 112.5 mV and achieved a time resolution of 3.6 ps (σ). Increasing the gain further would cause some loss of information at the zero crossings and degrade the resolution again.

Table 2 shows that board A gives a better time resolution for the PT-test than board C. Both boards show the expected period time of about 10 ns.

One can also see that TC N is about 15 times better than TC A and also provides the correct period time of 10 ns, even when TC N for channel# 1 is applied to channel# 2.

Additionally, it demonstrates that every channel has to be calibrated individually to obtain a time resolution below 10 ps.

### 2) The Split Pulse Test (SP-test)

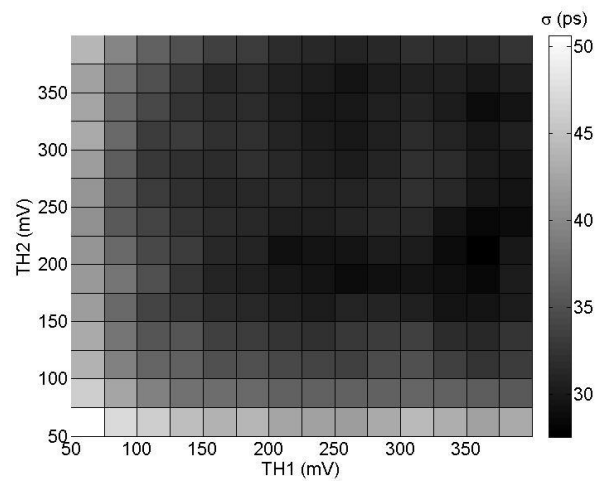

Fig. 10: SP-test (2 points) obtained with board C with 50 ns delay. The optimal TH settings of TH1 = 350 mV and TH2 = 200 mV give a time resolution of 28 ps (σ).

The results for the SP-test of a 50 ns delay are shown in Fig. 10 and Fig. 14. Fig. 10 shows the time resolution result of the SP-test for board C in dependence on the first threshold TH1 (first signal) and the second threshold TH2 (second signal) of the used digital leading edge (DLE) discriminator. The optimum time resolution is 28 ps (σ) with the signal shown in Fig. 7, which was digitized with 8 bit resolution and a full range of ~ 1.1V.

Fig. 11 illustrates results of a SP-test, where the time resolution in dependency of the delay is shown. Board D (large crosses) was running at 2 GSPS with a resolution of 10 bits. The two other boards were running at 5 GSPS using the DRS4 chip. The x-symbols mark the results of Board A using TC A and indicate a time resolution variation from 25 ps to 55 ps (σ). The same board achieves a time resolution of about 3 ps (σ) for all delays when TC N was applied (circles). The dots stand for the measurement points of board B using TC B. The 2-point-DLE-discriminator was used in all four cases.

Fig. 12 shows a zoom-in of the board A (TC N) curve from Fig. 11. For the same dataset two additional analysis methods are also shown. The circles and the dots were calculated with a DLE-discriminator. For the 50-point measurement (crosses) the cross-correlation method was used. One can see that the time resolution progresses when using more points. The time resolution curve shows approximately a linear rising behavior for this three cases.

| used board | used time calibration | $\frac{1}{SNR} = \frac{\Delta u}{U^*}$ ($*\approx 63\ mV$) | | Mean Value (ps) ± σ (ps) [expected σ with (1)] (ps) | |
|---|---|---|---|---|---|
| | | ch1 | ch2 | ch1 | ch2 |
| C | - | 0.0827 | 0.0702 | 10000.9 ±20.80[23] | 10000.0 ±16.70 [20] |
| C | - | 0.0163 | 0.0138 | 10000.1 ± 5.20[4.6] | 10000.0 ± 4.60 [4.0] |
| C | - | 0.0090 | 0.0076 | 10000.1 ± 3.90[2.5] | 10000.0 ± 3.60 [2.1] |
| A | TC N for each ch# | 0.0068 | 0.0071 | 10000.0 ± 3.11[1.9] | 10000.0 ± 3.23 [2.0] |
| A | TC N (always ch1) | 0.0068 | 0.0071 | 10000.0 ± 3.11[1.9] | 10000.0 ±13.20 [2.0] |
| A | TC A | 0.0068 | 0.0071 | 10004.5 ±48.05[1.9] | 10003.2 ±52.11 [2.0] |

Table 2: PT-test results for board A with 3 different TC & board C with 3 different gains. The theoretical best time resolution according to formula (1) is shown in brackets.

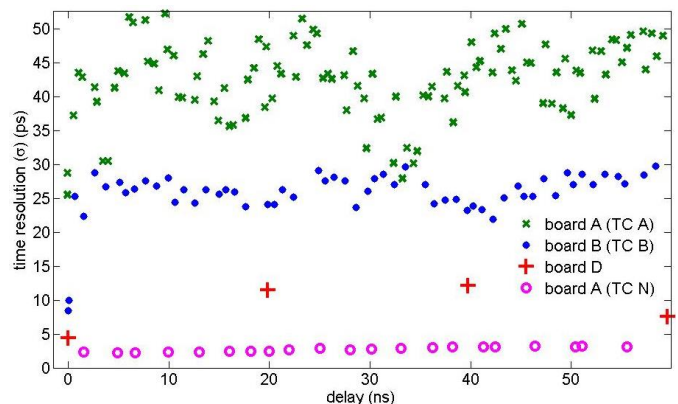

Fig. 11: SP-test measurements (2 points) obtained for 3 different boards.

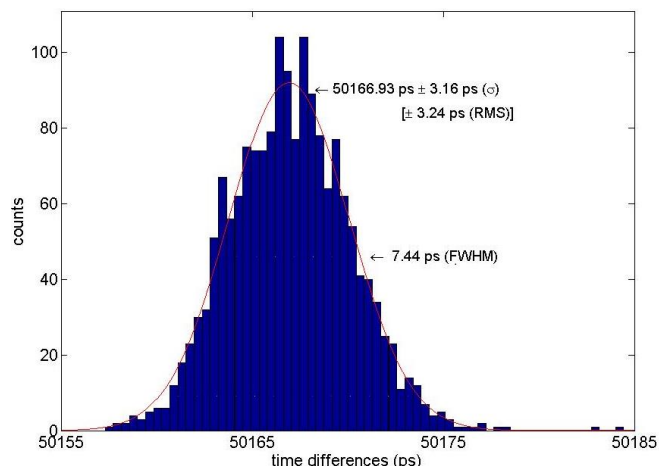

Fig. 13: SP-test distribution for the 2-point-DLE-discriminator (second to last triangle from Fig. 12).

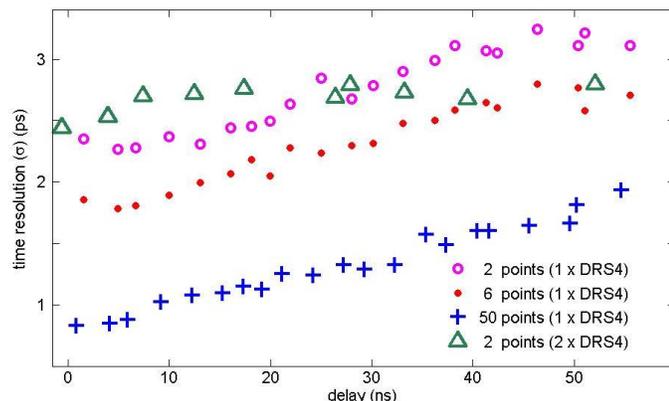

Fig. 12: SP-test measurements with board A after applying the TC N method. The triangles show the SP-test results of two independently running boards A, where a 2-point-DLE-discriminator was used. For the other 3 curves, where the SP-test was performed only with one board A, the same dataset was analyzed three times using different analyzing techniques.

The triangles in Fig. 12 demonstrate the time resolution results of the SP-test for two independently running boards A, synchronized by a dedicated split clock as described previously (Fig. 6b). The DLE-discriminator using only 2 points results in a time resolution better than 2.8 ps (σ) for all delays. The best achieved time resolution between two independently running DRS4 chips when using cross-correlation was better than 1.65 ps (σ) for any delay.

Fig. 13 demonstrates a single time resolution result of the SP-test from Fig. 12. A delay of about 50 ns was used and all measured 2000 events were plotted. One can see two outliers appearing at the right side.

The effect of using the wrong offset correction level with the DRS4 is illustrated in Fig. 14. The same experimental setup as shown in Fig. 7 is now digitized with 12 bit resolution. Additionally, we shifted intentionally the baseline by applying a 400 mV DC offset. The time resolution of board B as shown in Fig. 14 for a given TH of 300mV is 4.6 ps (σ) when using TC N. As expected the best TH2 level is lower than the optimal TH1 level, since the split signal is 15 % smaller in the second channel. When running the DRS4 of board B at the optimal DC offset level and applying the gain correction for board B one will result in 3ps (σ) as shown for board A in the 50 ns delay case of Fig. 11.

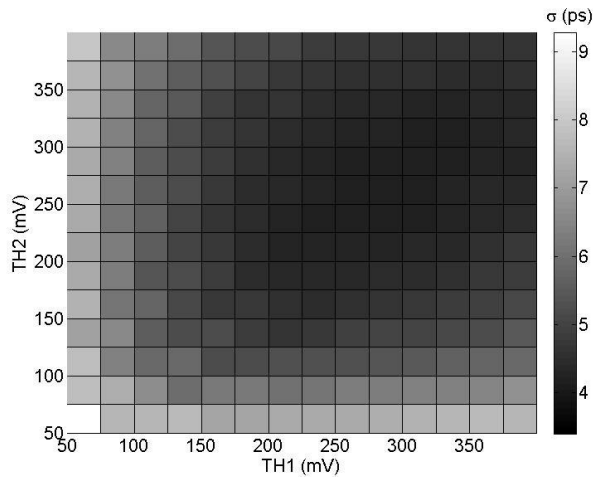

Fig. 14: SP-test (2 points) with the same delay as in Fig. 10 for board B. The optimal TH settings TH1 =275mV and TH2=250mV give a result of 4.4ps (σ).

### 3) The Coincident Resolving Time Test (CRT-test)

Fig. 15 shows two time resolution results of the CRT-test as a function of the thresholds TH1 (PMT# 1) and TH2 (PMT# 2) of the used 2-point-DLE discriminator. TH1 and TH2 are given in percentage of the average photo peak height of the particular channel. 65 ps CRT (σ) with TC B for best THs was measured and 74 ps CRT (σ) with the TC N. See Discussion for more details.

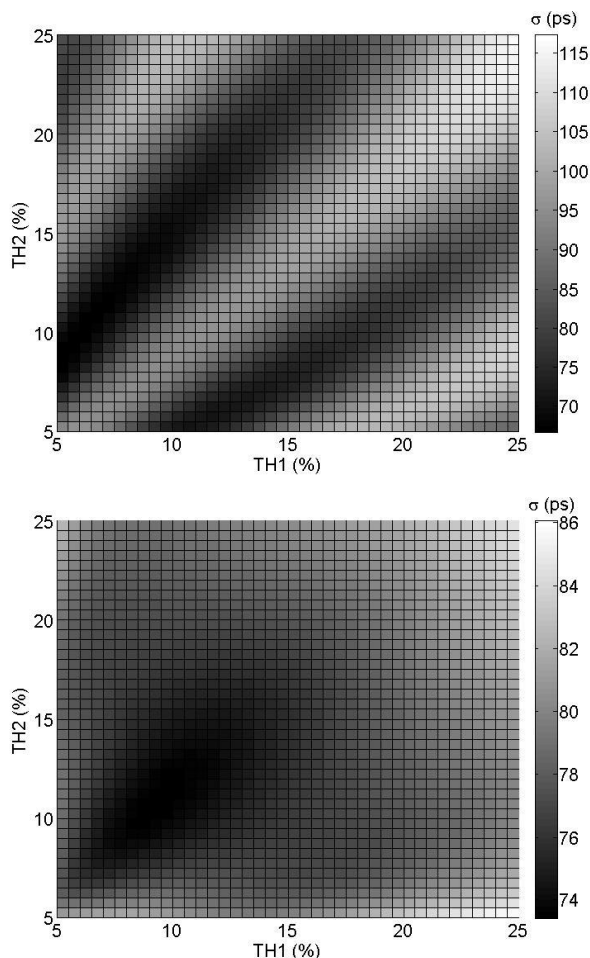

Fig. 15: Results for the CRT-test measured with board B. Two measurements with changing from TC B (top) to TC N (bottom) were performed.

### 4) The Temperature Time Dependence Test (TTD-test)

In Fig. 16 the temperature dependency of the time resolution of the DRS4 chip is plotted. One can see that half of the $\Delta t_{a,b}$ are varying less than 2 ps. However $\Delta t_{1,100}$ shows a maximum change of -12 ps compared to the 5°C case. The ten regions summed together result in no time change, indicating that the sampling speed remained stable. The reason for the temperature dependence is not completely clear, but can to some part be contributed to gradients of transistor parameters along the chip wafer.

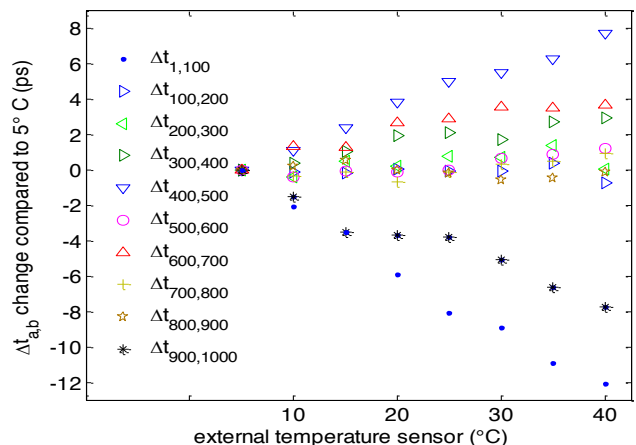

Fig. 16: TTD-test of board A. For each of the 8 measurements 1000 digitized waveforms were analyzed. Ten global time difference $\Delta t_{a,b}$ are showed in 20 ns steps. All $\Delta t_{a,b}$ values are around 20 ns and were subtracted from its corresponding 5°C case to illustrate the influence of temperature change for the DRS4. .

## VI. DISCUSSION

Previous TCs, like TC A and TC B, predicted not-alternating $\Delta t_i$ of 200 ps with an σ of 4 ps. By using TC N for the DRS4 chip we discovered that the true value of the sampling intervals $\Delta t_i$ alternate between 130 ps and 270 ps with a σ of ~23 ps at 5 GSPS (Fig. 9). The reason for these alternations lies in the layout of the DRS4 chip. The 1024 sampling cells of one channel are not linearly arranged, but folded due to the limited size of the die. This causes odd and even cells to see a different environment on the chip and to be differently connected to the power rails. At lower sampling speeds one will find a similar alternating behavior of the DRS4 chip. Since the signal edges between the inverters have longer rise times at lower sampling speed, internal noise plays a bigger effect and the timing performance degrades about inversely with the sampling frequency $\Delta t \sim 1/f_s$.

### 1) The Period Time Test (PT-test)

The PT-Test in Table 2 shows that TC N for board A yields in the expected period time of 10 ns and a time resolution of 3.1 ps. With the best gain, board C only reaches 3.6 ps. Since the time resolution is proportional to the rise time of the signal at a given signal-to-noise, a higher gain setting for board C results in a better time resolution, although the peaks of the test-signal will then be clipped.

The channel-by-channel TC (illustrated in Table 2) can be understood by looking at Fig. 1. Every analog switch connected to the sampling capacitors has a separate buffer. The transition times of these buffers are different due to the above-mentioned variations in chip process parameters.

Board A had issues before with measuring long time differences ([20] and Fig. 11). This problem is now understood and has several reasons. The reasons will be addressed in decreasing significance. First, the TC frequency that was used for TC A had a 120 ps time jitter with a non-Gaussian distribution. Second, the true sampling intervals alternate and do not show a 200 ps ± 4ps behavior as predicted by TC A & B, but a 200 ps ± 74 ps behavior. Third, for every channel of an SCA chip individually TC is mandatory. Fourthly, the true sampling speed differed from the expected sampling speed by about 0.01 %. The measurement of the true sampling speed with TC N and the usage of a more precise oscillator with less jitter on version 5 of board A solved this problem. Fifthly, a 2.5 MHz digital signal on board A on a PCB trace close to the DRS4 chip induced some instability of the PLL inside the DRS4 which lead to two distinguished alternating sampling speeds. A redesign of the PCB with better shielding of this signal fixed that problem.

*2) The Split Pulse Test (SP-test)*

Looking at the 50 ns delay result from Fig. 12 compared to Fig. 14 shows a time resolution degradation from 2.9 ps to 4.6 ps for the same measurement when using a different offset level and no gain correction. We used a DLE-discriminator where the stability of the baseline is mandatory. This underscores that the cell-to-cell gain spread leading to level-dependent offsets up to 0.5 mV (see data sheet [21], Plot 2) has a considerable effect on the time resolution. A cell-by-cell calibration of the non-linearity could therefore improve the time resolution even further.

Under similar conditions (Fig. 7) board C achieved 28 ps(Fig. 10). This was predictable since it only uses 8 bits for a full range of ~1.1V. The best time resolution for board C performing the SP-test was around 8 ps, when the gain was increased by a factor of 10 to a full range of 0.1125V, but with the penalty of above-mentioned clipping.

Fig. 11 shows the SP-test for different boards and TCs. In comparison to TC B the TC A on board A is about two times worse. This is mainly caused by the TC-signal that was used for the TC A, which has a jitter greater than 100 ps. Also, the time resolution should remain constant when increasing the delay, which is not the case for TC A or TC B. Instead, the DRS4 digitizers show a strong correlation between the delay and its corresponding time resolution behavior when using TC A or TC B.

The large DRS4 time resolution improvement when comparing TC N with the other TCs can be explained with the following: when looking at the time axes of an uncalibrated channel running at 5 GSPS, the maximal error is around 800 ps. When comparing TC N with TC A or TC B, the maximum error is still around 150 ps. This is because the other TCs provide almost equidistant sampling intervals of 200 ps with a σ of 4 ps and not the alternating behavior mention at the beginning of section VI. However, one can also see in Fig. 11 that after using TC N, the curve for board A lies below the curve of board D as expected, because the DRS4 was running at 5 GSPS using 12 bits and board D at 2 GSPGS using 10 bits.

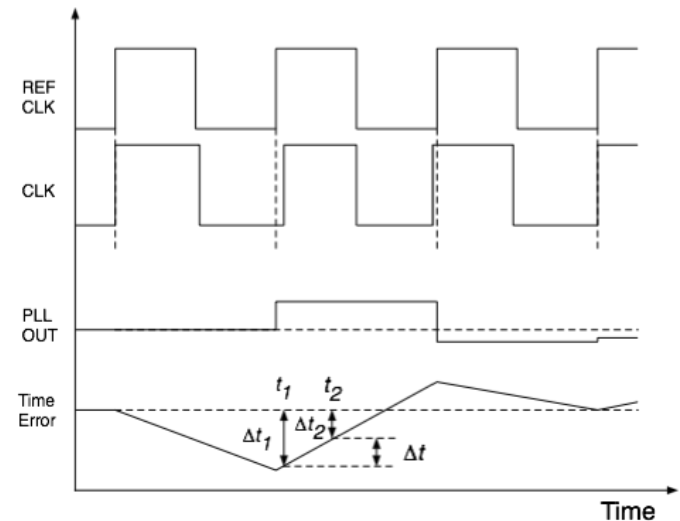

Fig. 17: PLL phase jitter in an SCA chip. REFCLK is the external (exact) reference clock, CLK the frequency of the inverter chain, PLLOUT is the control voltage from the PLL and Time Error the deviation of the sampling time from the exact time.

In Fig. 12 one can see that the time resolution is improved by using more sampling points of the signal, as expected. We observe however even for short delays an improvement by less than $1/\sqrt{n}$, which has three reasons. Firstly, the 2-point measurement used the samples with the best SNR (the highest slope of the signal), so the other sampling points will contribute less to the improvement of the measurement. Secondly, the measured samples are not statistically independent as required by the $1/\sqrt{n}$ law. This comes from the fact that the noise spectrum of the measured signal has slower components, which can affect adjacent samples in a coherent way. Thirdly, the linear interpolation is not the optimal fitting method.

The increase of the time resolution with the cable delay comes from the fact that the sampling speed varies around its nominal value due to the residual phase jitter of the PLL in the DRS4 chip. This can be seen in Fig. 16 which shows the deviation of the sampling time from the exact time due to the PLL time jitter. If two signals are sampled at times $t_1$ and $t_2$ on separate channels driven by the same inverter chain, their deviation from the perfect time is $\Delta t_1$ and $\Delta t_2$, respectively. The relative time error between the two signals is $\Delta t$, which is proportional to the time distance $t_2 - t_1$ as long as the time difference is smaller than the clock period. Since the inverter chain in case of the DRS4 is 1024 cells long, the time resolution in Fig. 13 increases from 0.8 ps at 0 ns delay (which is close to the theoretical optimum) to 4.8 ps at 200 ns delay (extrapolated), reflecting the PLL jitter of about 4 ps.

Fig. 12 also shows that this increase of the time resolution with the cable delay can be compensated by the additional sampled 100 MHz clock information that was used to synchronize two DRS4 chips. As expected the time resolution of the SP-test is worse compared to the single DRS4 case for short delays. On the other hand, cable delays above 25 ns already result in better time resolution and were measured for cable delays up to 150 ns to be less than 2.8 ps ($\sigma$).

### 3) The Coincident Resolving Time Test (CRT-test)

In Fig. 15 the importance of applying TC N is shown with an example of a real PET measurement (Fig. 8). Looking at measurements of board B using TC B one will get a wrong result. This is illustrated in the top figure of Fig. 15, which shows 3 minima instead of just one expected and is 7 ps better CRT ($\sigma$) than possible. We know that the time resolution result of board B (TC B) is wrong because we double checked the experiment with an oscilloscope with 20 GSPS and an adjusted gain for a comparable 12 bit resolution. Although we digitized 4 times faster, we only archived a CRT ($\sigma$) of 72 ps for best threshold (TH) settings.

The reason for the wrong result is the following:

We know that the effective sampling interval $\Delta t_i$ of TC B provides almost equidistant 200 ps sampling intervals. As mentioned above the true $\Delta t_i$ alternates between 130 ps and 270 ps. When changing the TH settings as shown Fig. 15 one will also measure different time differences between the two PET signals (walk effect). The three minima also represent regions of similar time differences. When this time difference is calculated by interpolating between two neighboring cells that are mainly 270 ps apart but wrongly considered to be 200 ps, the calculated $\sigma$ will be smaller than in reality and will therefore result in a wrongly considered as better time resolution result. Considering sampling intervals of around 130 ps one will also find regions with bigger time resolution than possible as shown in Fig. 15.

### 4) The Temperature Time Dependence Test (TTD-test)

The temperature stability test of board A in Fig. 16 shows that the variation of $\Delta t_{a,b}$ with the temperature is less than 1-2 ps in a temperature range from 5°C to 10°C. However, the variation of $\Delta t_{a,b}$ cannot be ignored for bigger temperature variations. $\Delta t_{200,700}$ can be predicted by summing the corresponding $\Delta t_{a,b}$ from Fig. 16 and results in an expected change of 23 ps for $\Delta t_{200,700}$ when increasing the temperature by 35° for temperature changes around 1-2 °C the $\Delta t_{a,b}$ variation is below the extrapolated PLL jitter of around 2 ps for this delay. Thus, a temperature adjusted TC is mandatory if an excellent time resolution is needed and if the temperature varies more than 2°C. However, the TC was tested to be valid over several months. Also the sampling frequency stays the same for the tested temperatures as shown in Fig. 16 when adding all 10 temperature points together.

## VII. Conclusion

The novel TC N gives excellent results for DRS4-based time measurements. Since the limitations of time measurements are very similar in most SCA chips, such as unequal propagation times of inverter chains and buffers, it is very likely that this calibration is well applicable also for other SCA chips.

In the DRS4 case a time resolution improvement by a factor of 8 to 15 has been achieved (Fig. 11). The performance is now much better compared to an oscilloscope, while the costs of an SCA-based system are one order of magnitude lower. For a single DRS4, up to 30 ns delay (Fig. 12), the SP-test using cross-correlation is below 1.4 ps ($\sigma$), giving a single time resolution better than $1.38 \text{ ps}/\sqrt{2} = 0.98 \text{ ps}$ ($\sigma$). However, when performing the SP-test on two independently running DRS4 chips and using a simple 2-point-DLE-discriminator, a single time resolution better than $2.80 \text{ ps}/\sqrt{2} = 1.98 \text{ ps}$ ($\sigma$) was achieved for any cable delay.

Thus, the DRS4 provides an excellent measurement platform for applications in particle physics or in PET medical imaging.

## Acknowledgment

The authors would like to thank the colleagues from Siemens Medical Imaging, especially Matthias Schmand, Nan Zhang, Robert Mintzer, Sanghee Cho, Peter Cohen and Larry Byars for supporting the work with the DRS4.

We are also grateful to Ueli Hartmann, Christoph Parl, Chih-Chieh Liu, Frederic Mantlik, Armin Kolb, Mathew Divine and Jeanine Adam for helpful conversations and/or support.

We would like to thank the University Hospital Tübingen for making it possible to file in an international patent application (No. PCT/EP2013/070892) containing several TCs including the demonstrated new TC method.

## References

[1] S. A. Kleinfelder, W. C. Carithers, R. P. Ely, C. Haber, F. Kirsten, and H. G. Spieler, "A flexible 128 channel silicon strip detector instrumentation integrated circuit with sparse data readout," *IEEE Trans. Nucl. Sci.*, vol. 35, no. 1, pp. 171–175, Feb. 1988.

[2] G. M. Haller and B. A. Wooley, "An analog memory integrated circuit for waveform sampling up to 900 MHz," *IEEE Trans. Nucl. Sci.*, vol. 41, no. 4, pp. 1203–1207, 1994.

[3] E. Delagnes, Y. Degerli, P. Goret, P. Nayman, F. Toussenel, and P. Vincent, "SAM: A new GHz sampling ASIC for the H.E.S.S.-II front-end electronics," *Nucl. Instruments Methods Phys. Res. Sect. A Accel. Spectrometers, Detect. Assoc. Equip.*, vol. 567, no. 1, pp. 21–26, Nov. 2006.

[4] G. S. Varner, L. L. Ruckman, J. W. Nam, R. J. Nichol, J. Cao, P. W. Gorham, and M. Wilcox, "The large analog bandwidth recorder and digitizer with ordered readout (LABRADOR) ASIC," *Nucl. Instruments Methods Phys. Res. Sect. A Accel. Spectrometers, Detect. Assoc. Equip.*, vol. 583, no. 2–3, pp. 447–460, Dec. 2007.

[5] S. Ritt, R. Dinapoli, and U. Hartmann, "Application of the DRS chip for fast waveform digitizing," *Nucl. Instruments Methods Phys. Res. Sect. A Accel. Spectrometers, Detect. Assoc. Equip.*, vol. 623, no. 1, pp. 486–488, Nov. 2010.

[6] E. Oberla, J.-F. Genat, H. Grabas, H. Frisch, K. Nishimura, and G. Varner, "A 15GSa/s, 1.5GHz bandwidth waveform digitizing ASIC," *Nucl. Instruments Methods Phys. Res. Sect. A Accel. Spectrometers, Detect. Assoc. Equip.*, vol. 735, pp. 452–461, Jan. 2014.

[7] J. Adam, et al. , *The MEG detector for $\mu^+ \to e^+ \gamma$ decay search*, vol. 73, no. 4. 2013, p. 2365.

[8]  J. Sitarek, M. Gaug, D. Mazin, R. Paoletti, and D. Tescaro, "Analysis techniques and performance of the Domino Ring Sampler version 4 based readout for the MAGIC telescopes," *Nucl. Instruments Methods Phys. Res. Sect. A Accel. Spectrometers, Detect. Assoc. Equip.*, vol. 723, pp. 109–120, Sep. 2013.

[9]  D. Breton, E. Delagnes, J. Maalmi, K. Nishimura, L. L. Ruckman, G. Varner, and J. Va'vra, "High resolution photon timing with MCP-PMTs: A comparison of a commercial constant fraction discriminator (CFD) with the ASIC-based waveform digitizers TARGET and WaveCatcher," in *IEEE Nuclear Science Symposuim & Medical Imaging Conference*, 2010, pp. 856–864.

[10]  J. A. Aguilar, et al., "Performance of the front-end electronics of the ANTARES neutrino telescope," *Nucl. Instruments Methods Phys. Res. Sect. A Accel. Spectrometers, Detect. Assoc. Equip.*, vol. 622, no. 1, pp. 59–73, Oct. 2010.

[11]  B. W. Jakoby, Y. Bercier, M. Conti, M. E. Casey, B. Bendriem, and D. W. Townsend, "Physical and clinical performance of the mCT time-of-flight PET/CT scanner.," *Phys. Med. Biol.*, vol. 56, no. 8, pp. 2375–89, Apr. 2011.

[12]  D. R. Schaart, S. Seifert, R. Vinke, H. T. van Dam, P. Dendooven, H. Löhner, and F. J. Beekman, "LaBr(3):Ce and SiPMs for time-of-flight PET: achieving 100 ps coincidence resolving time.," *Phys. Med. Biol.*, vol. 55, no. 7, pp. N179–89, Apr. 2010.

[13]  D. Breton, E. Delagnes, J. Maalmi, K. Nishimura, L. L. Ruckman, G. Varner, and J. Va'vra, "High resolution photon timing with MCP-PMTs: A comparison of a commercial constant fraction discriminator (CFD) with the ASIC-based waveform digitizers

TARGET and WaveCatcher," *Nucl. Instruments Methods Phys. Res. Sect. A Accel. Spectrometers, Detect. Assoc. Equip.*, vol. 629, no. 1, pp. 123–132, Feb. 2011.

[14]  A. M. Makankin, V. V. Myalkovskiy, V. D. Peshekhonov, S. Ritt, and S. E. Vasilyev, "A direct time measurement technique for the two-dimensional precision coordinate detectors based on thin-walled drift tubes," *Nucl. Instruments Methods Phys. Res. Sect. A Accel. Spectrometers, Detect. Assoc. Equip.*, vol. 735, pp. 649–654, Jan. 2014.

[15]  J.-F. Genat, *private communication*. .

[16]  PSI, *DRS4 Evaluation Board User's Manual Rev 2, Rev 3 (board A), Rev 4 or Rev 5*. http://www.psi.ch.

[17]  CAEN, *V1742 Technical Information Manual Rev 6*. http://www.caen.it.

[18]  K. Nishimura and A. Romero-Wolf, "A Correlation-based Timing Calibration and Diagnostic Technique for Fast Digitizing ASICs," *Phys. Procedia*, vol. 37, pp. 1707–1714, 2012.

[19]  J. Wang, L. Zhao, C. Feng, S. Liu, and Q. An, "Evaluation of a Fast Pulse Sampling Module With Switched-Capacitor Arrays," *IEEE Trans. Nucl. Sci.*, vol. 59, no. 5, pp. 2435–2443, Oct. 2012.

[20]  A. Ronzhin, et al., "Waveform digitization for high resolution timing detectors with silicon photomultipliers," *Nucl. Instruments Methods Phys. Res. Sect. A Accel. Spectrometers, Detect. Assoc. Equip.*, vol. 668, pp. 94–97, Mar. 2012.

[21]  PSI, *DRS4 data sheet*. http://www.psi.ch.